\documentclass[referee,sn-basic]{sn-jnl}

\usepackage{graphicx}%
\usepackage{multirow}%
\usepackage{amsmath,amssymb,amsfonts}%
\usepackage{amsthm}%
\usepackage{mathrsfs}%
\usepackage[title]{appendix}%
\usepackage{xcolor}%
\usepackage{textcomp}%
\usepackage{manyfoot}%
\usepackage{booktabs}%
\usepackage{algorithm}%
\usepackage{algorithmicx}%
\usepackage{algpseudocode}%
\usepackage{listings}%
\usepackage{algorithm,algpseudocode,amsmath,amssymb,amsthm,amsfonts,blkarray,bbm,bm,color,setspace,graphics,graphicx,indentfirst,url,caption,graphicx,multicol,multirow,longtable,tabu,tikz,verbatim,enumerate,footnote,multirow,mathtools,pifont,xcolor,url,caption,subcaption}

\theoremstyle{thmstyleone}%
\theoremstyle{thmstyletwo}%

\theoremstyle{thmstylethree}%

\begin{document}

\title[Temporal clustering of functional trajectories]{A Bayesian Nonparametric Approach for Clustering Functional Trajectories over Time}


\author[1]{Mingrui Liang}
\author[2]{Matthew D. Koslovsky}
\author[3]{Emily T. H\'{ebert}}
\author[4]{Darla E. Kendzor}
\author*[1]{Marina Vannucci}\email{marina@rice.edu}

\affil[1]{\orgdiv{Department of Statistics}, \orgname{Rice University}, \orgaddress{\state{TX}, \country{USA}}}
\affil[2]{\orgdiv{Department of Statistics}, \orgname{Colorado State University}, \orgaddress{\state{CO}, \country{USA}}}
\affil[3]{\orgdiv{Department of Health Promotion and Behavioral Sciences}, \orgname{UT School of Public Health}, \orgaddress{\state{TX}, \country{USA}}}
\affil[4]{\orgname{University of Oklahoma Health Sciences Center}, \orgaddress{\state{OK}, \country{USA}}}


\abstract{Functional concurrent, or varying-coefficient, regression models are commonly used in biomedical and clinical settings to investigate how the relation between an outcome and observed covariate varies as a function of another covariate. In this work, we propose a Bayesian nonparametric approach to investigate how clusters of these functional relations evolve over time. Our model clusters individual functional trajectories within and across time periods while flexibly accommodating the evolution of the partitions across time periods with covariates. Motivated by mobile health data collected in a novel, smartphone-based smoking cessation intervention study, we demonstrate how our proposed method can simultaneously cluster functional trajectories, accommodate temporal dependence, and provide insights into the transitions between functional clusters over time.}

\keywords{Clustering; Functional data analysis; Hierarchical Dirichlet Process; Nonparametric Bayes;  Smoking Cessation; Temporal Dependence.}



\maketitle
\section{Introduction}
\label{s:intro}

Functional concurrent, or varying-coefficient, regression models are commonly used in biomedical and clinical settings to investigate how the relation between an outcome and observed covariate varies as a function of another covariate \citep{tan2012time, kim2018additive,leroux2018dynamic}. These techniques belong to the general class of functional data analysis (FDA) methods and are a special case of function-on-function regression models in which the functional response and functional covariates are collected concurrently\citep{morris2015functional,reiss2017methods,zhang2011functional,maity2017nonparametric}. For example, this work is motivated by functional data collected in a mobile health (mHealth) study in which participants were repeatedly prompted with ecological momentary assessments (EMAs) to generate near real-time information regarding the dynamic relation between potential risk factors and smoking behaviors over time.  

A common objective in FDA research is to cluster groups of observations that share similar functional trends to identify homogeneous subpopulations \citep{tarpey2003clustering, ferreira2009comparison,abraham2003unsupervised,james2003clustering}. 
Existing approaches for functional clustering in the frequentist setting include distance-based methods, similar to the k-means algorithm \citep{abraham2003unsupervised}, or model-based methods \citep{james2003clustering}.
Bayesian model-based functional clustering methods are also widely used since they allow for the incorporation of prior knowledge about functional trajectories and provide a framework for uncertainty quantification. 
Additionally, Bayesian methods offer an advantage by relaxing assumptions regarding the number of clusters in the data through the use of nonparametric prior specifications, such as the Dirichlet process (DP) \citep{ferguson1973bayesian}, hierarchical DP process \citep{Teh2006hdp}, and Pitman-Yor process \citep{pitman1997two}. These processes are frequently used to govern clustering allocation due to their computational simplicity in various research settings \citep{suarez2016bayesian,white2020multivariate,das2021modeling, koslovsky2024dynamic}.

Our interest is in clustering functional trajectories over time, an aspect of temporal, i.e., time-dependent, clustering.
In recent years, Bayesian nonparametric prior formulations that accommodate temporal dependence between cluster allocations have emerged \citep{nieto2012time,gutierrez2016time,jo2017dependent}. Most of these developments introduce temporal dependence across the cluster allocations through the weights and/or atoms of the infinite mixture using common time-series techniques.
While these methods offer several advantages \citep{ascolani2021predictive}, the cluster evolution is only implied in the process and the sequence of cluster allocations is not explicitly modeled,  limiting inference regarding how clusters transition over time. More direct approaches for modeling the sequence of cluster allocations have also been proposed \citep{caron2007generalized, caron2017generalized,zanini2019bayesian}. However they typically only exhibit  weak temporal dependence between clusters, do not model element-level transitions, and/or make restrictive assumptions on the number of time points or clusters. \cite{page2021dependent} address these limitations by directly modeling the cluster transitions individually for each element. Their approach not only allows for a more intuitive dependence pattern between cluster allocations over time, but also conveniently enables individual-level modeling and inference in temporal dependent clustering problems. 
 
In this work, we introduce a novel Bayesian temporal clustering method that accommodates functional response and covariate data that are collected at the same time, which, to our knowledge, has not been addressed with existing methods. Our method clusters individual functional trajectories within a given time period and accommodates the evolution of the partitions across time periods with covariates.
Additionally, our proposed method facilitates the sharing of similar trajectories both within and across time periods via hierarchical nonparametric priors, which provides an extra layer of dependence structure. 
We apply our model to intensive longitudinal mHealth data collected from a smartphone-based smoking cessation intervention study, referred to as the PREVAIL II study \citep{Kendzor2023plus}. During the study, participants were repeatedly prompted with  EMAs of their current environment, affect, behaviors, social interactions, and recent smoking behaviors on a study-provided smartphone over a five week period (one week before and four weeks after the quit attempt).  Participants' smoking status was biochemically verified each week during the quit attempt, and they were given the option of attending additional counseling sessions at each visit.  
By using our proposed model, we can explore the temporal evolution of individual smoking trajectories week to week  and identify subgroups of participants with similar behavioral trajectories of smoking within each week. This information can help clinicians prioritize and tailor treatment to individuals based on their risk level and treatment needs.

The rest of the paper is organized as follows. In section 2, we introduce our Bayesian temporal functional clustering method. In section 3, we evaluate and compare the clustering and estimation performance of our proposed method on simulated data and perform a sensitivity analysis. In section 4, we apply our model to smoking cessation data collected in the PREVAIL II study. In section 5, we provide concluding remarks. 

\section{Hierarchical Temporal Dependent Functional Clustering}
\label{s:model}

\subsection{Functional Concurrent Regression Model}
Consider a study with $i=1,\dots,N$ participants, where each participant is observed for an equal number of time periods $j=1,\dots,J$. 
In our motivating application, the $J$ periods correspond to the week-long observation periods after a participant has attended their weekly verification and counseling sessions (i.e., week 1, week 2, ..., week $J$ after the quit attempt). We assume that all participants have at least one observed response within each of the $J$ periods.

Let $Y_{ij}(\cdot) \in \{0,1\}$ be the functional binary response (e.g., momentary smoking status), which is modeled by a  functional concurrent regression model:
\begin{equation}
\label{eq:logit}
    \text{logit}(P(Y_{ij}(t_{ijm})=1)) = \mathcal{B}_{ij}(t_{ijm}) + \bm{Z}_{ij} \bm\theta_j,
\end{equation}
where $t_{ijm}$ is the time point where each observation occurs, $m=1,\dots,M_{ij}$, and $M_{ij}$ is the total number of observations for the $i^{th}$ participant in the $j^{th}$ period. Note that for each participant $i$ within each period $j$, the number of observations could potentially be different. 
$\mathcal{B}_{ij}(t_{ijm})$ represents a participant-specific smooth function in period $j$,   $\bm{Z}_{ij}$ is a set of observed baseline risk factors (e.g., weight, age, etc.), and $\bm\theta_j \sim \text{N}(\bm 0, \Sigma_\theta)$ is the corresponding vector of regression coefficients that remain constant within a given period. 

We approximate the smooth functions using cubic splines, denoted as $\mathcal{B}_{ij}(\cdot) = \bm{\mathcal{T}}_{ij} \bm{\beta}_{ij}$, with $\bm{\mathcal{T}}_{ij}$ an $M_{ij}\times Q$-dimensional matrix of basis functions and $\bm{\beta}_{ij}$ the corresponding $Q$-dimensional vector of spline coefficients. To avoid overfitting the estimated smooth function, we penalize each coefficient with a horseshoe prior \citep{carvalho2010horseshoe} 
\begin{eqnarray}
\beta_{ijq} \sim N(0,\tau_{ij}^2 \lambda_{ijq}^2),
\label{horseshoe}
\end{eqnarray}
for $q=1,\dots,Q$, with $\tau_{ij}^2$ and $\lambda_{ijq}^2$ representing the global and local variance terms, respectively.
We assume 
$\tau_{ij}^2|\nu_{\tau_{ij}} \sim \mathcal{IG}(1/2,1/\nu_{\tau_{ij}})$,
$\lambda_{ijq}^2|\nu_{\lambda_{ijq}} \sim \mathcal{IG}(1/2,1/\nu_{\lambda_{ijq}})$
and $\nu_{\lambda_{ij1}},\dots,\nu_{\lambda_{ijQ}},\nu_{\tau_{ij}} \sim \mathcal{IG}(1/2,1)$, where $\nu$ is an auxiliary parameter introduced for efficient sampling, following \cite{makalic2015simple}, and $\mathcal{IG}$ represents an Inverse-Gamma distribution.
In addition to its sampling efficiency, the horseshoe prior offers several other advantages compared to alternatives such as Gaussian or Laplacian priors, including a balance between shrinkage and sparsity, adaptivity to signal strength, and robustness to outliers. Further, this approach is in general computationally more efficient than using Gaussian process priors \citep{petrone2009hybrid, scarpa2009bayesian} or other alternative approaches for modeling smooth functions.

\subsection{Temporal Random Partition Model for Functional Data}
\label{sec:spline}

In this study, we are interested in identifying similar patterns of participants' smoking behaviors after weekly counseling sessions as well as how these patterns evolve over the study window. We achieve this by placing a nonparametric prior on the spline coefficients used to approximate the functional trajectories. Our approach is motivated by the temporal random partition model (tRPM) introduced in \cite{page2021dependent}, which embeds temporal dependence among the clustering sequence. We extend the tRPM to accommodate functional data and to provide additional dependence across cluster allocations at different periods with a hierarchical structure. 

The idea behind the tRPM is that cluster allocations should change less (more) between highly (less) dependent partitions across periods. At period $j$, we assume the partition for all participants $i$ is represented as $\rho_j = \{S_{j1},\dots,S_{j{C_j}}\}$, where $C_j$ indicates the total number of within-period clusters for period $j$. We then assume a first-order Markovian structure among the clustering sequence (i.e., the partition at period $j$, $\rho_j$, only depends on the previous partition, $\rho_{j-1} = \{S_{j-1,1},\dots,S_{j-1,{C_{j-1}}}\}$). The joint distribution of $\bm{\rho}$ and $\bm{\gamma}$ is defined as
\begin{equation}    p(\bm{\gamma}_1,\rho_1,\dots,\bm{\gamma}_J,\rho_J)=p(\rho_J|\bm{\gamma}_J,\rho_{J-1})p(\bm{\gamma}_J)\cdots p(\rho_2|\bm{\gamma}_2,\rho_1)p(\bm{\gamma}_2)p(\rho_1).
\end{equation}
For each period $j$, we assume that   partition $\rho_j$ is induced by a Dirichlet process
\begin{equation}
\label{eq:dp}
\begin{aligned}
    \bm\beta_{ij}|G_j &\stackrel{\text{iid}}{\sim} G_j\\
    G_j|\alpha, G &\stackrel{\text{iid}}{\sim} \text{DP}(\alpha, G),
\end{aligned}
\end{equation}
where $\alpha$ represents the concentration parameter and $G$ the base distribution. 
To model the  cluster transitions between periods, we embed participant- and period-specific auxiliary parameters $\gamma_{ij} \in \{0,1\}$ that control the similarity between $\rho_{j-1}$ and $\rho_{j}$. For $j>1$, the ``fixed'' participants (i.e., $\gamma_{ij}=1, \forall i $) remain at the previous partition, and the ``flexible'' participants (i.e., $\gamma_{ij}=0, \forall i $) can change their cluster assignments. Therefore, only a subset of the participants will change their cluster assignments between periods. Intuitively, it is necessary to have compatible $\rho_{j-1}$ and $\rho_j$ in the evolution of the partitions, meaning that the reduced partition with only the ``fixed'' participants in periods $j-1$ and $j$ should remain clustered together. This prevents contradictory situations in which a participant is not allowed to join a different cluster (i.e., $\gamma_{ij}=1$), but the cluster allocation indicates otherwise. 
To model this assumption, let $R_j=\{i:\gamma_{ij}=1\}$ denote the fixed participants from $j-1$ to $j$, and  $\rho_j^{R}$ represent a reduced partition at period $j$ with only participants in set $R$. We then require that $\rho_{j-1}^{R_j}=\rho_j^{R_j}$ for all $j=2,\cdots,J$. We assume  $\gamma_{ij}\sim \text{Bernoulli}(\phi_{ij})$, where $\phi_{ij}$ controls the strength of the temporal dependence across periods. We provide additional flexibility in estimating the temporal dependence by modeling $\phi_{ij}$ as a logit link function for a $d_x$-dimensional vector of covariates, $\text{logit}(\phi_{ij}) = \bm{X}_{ij}\bm{\eta}_j$, at each time period $j=2,\dots,J$.  We note that in this context covariates $\bm X$ are used to account for temporal dependence, while covariates $\bm Z$ in Equation \eqref{eq:logit} are included to accommodate participant- and period-specific baseline effects. Under this parameterization, it is possible for $\bm X$ and $\bm Z$ to contain similar sets of covariates.
Lastly, we let $\bm{\eta}_j \sim N(\bm\mu_\eta,\bm\Sigma_\eta)$, where $\bm\mu_\eta$ and $\bm\Sigma_\eta$ are hyperparameters.

\subsection{Hierarchical tRPM}

The partition introduced above does not allow sharing of similar functional trajectories across different periods.
In other words, unique realizations of $\bm\beta_{ij}$ would only be shared within the $j^{th}$ period.  In practice, researchers may be interested in behavioral patterns that are shared within- and between-periods. Additionally, borrowing strength across periods can help improve estimation and simplify inference. 

To allow for clusters of behavioral trajectories to be shared across periods, we propose a hierarchical temporal dependent functional clustering approach, analogous to the hierarchical Dirichlet Process (HDP) \citep{Teh2006hdp} extension to the DP, which connects the spline coefficients $\bm\beta_{ij}$ at different periods by further assuming that the base distribution $G$ in equation \eqref{eq:dp} follows another DP,
\begin{equation}
\label{eq:hdp}
\begin{aligned}
G|\alpha_0, H \sim \text{DP}(\alpha_0, H), 
\end{aligned}
\end{equation} 
where $\alpha_0$ is a concentration parameter and $H$ is a base distribution. 
We denote this construction as htRPM$(\phi,\alpha, \alpha_0)$. Under this framework, we assume that there exists a total of $D$ unique realizations of $\bm\beta_{ij}$ across periods, and the unique realizations $\bm\beta_d^*$, $d = 1,\dots,D$, are sampled from $H$ to create a discrete mixture distribution $G$ for all periods $j$.
Within each period, mixture distributions $G_j$ are formed with the same elements $\bm\beta_d^*$ but with period-specific weights. Each $\bm\beta_{ij}$ is then sampled from $G_j$.
As a result, similar trajectories across different periods can share the same spline coefficients,
i.e., unique realizations $\bm\beta_d^*$ can be shared by participants across periods $j$, which naturally provides another layer of dependence across cluster allocations at different periods.
Note that within each time period, we may or may not observe all $D$ unique realizations $\bm\beta_d^*$. In other words, the number of clusters might differ for different time periods. 
Lastly, hyperparameters corresponding to the same unique realization of spline coefficients would be identical, that is $\tau_{ij} = \tau_{i'j'} = \tau^*_d$ if $\bm\beta_{ij} = \bm\beta_{i'j'} = \bm\beta^*_d, \forall i,j$. This similarly applies to $\bm\lambda$ and $\bm\nu$.

\begin{figure}[htb!]
\centering
\includegraphics[width=0.8\textwidth]{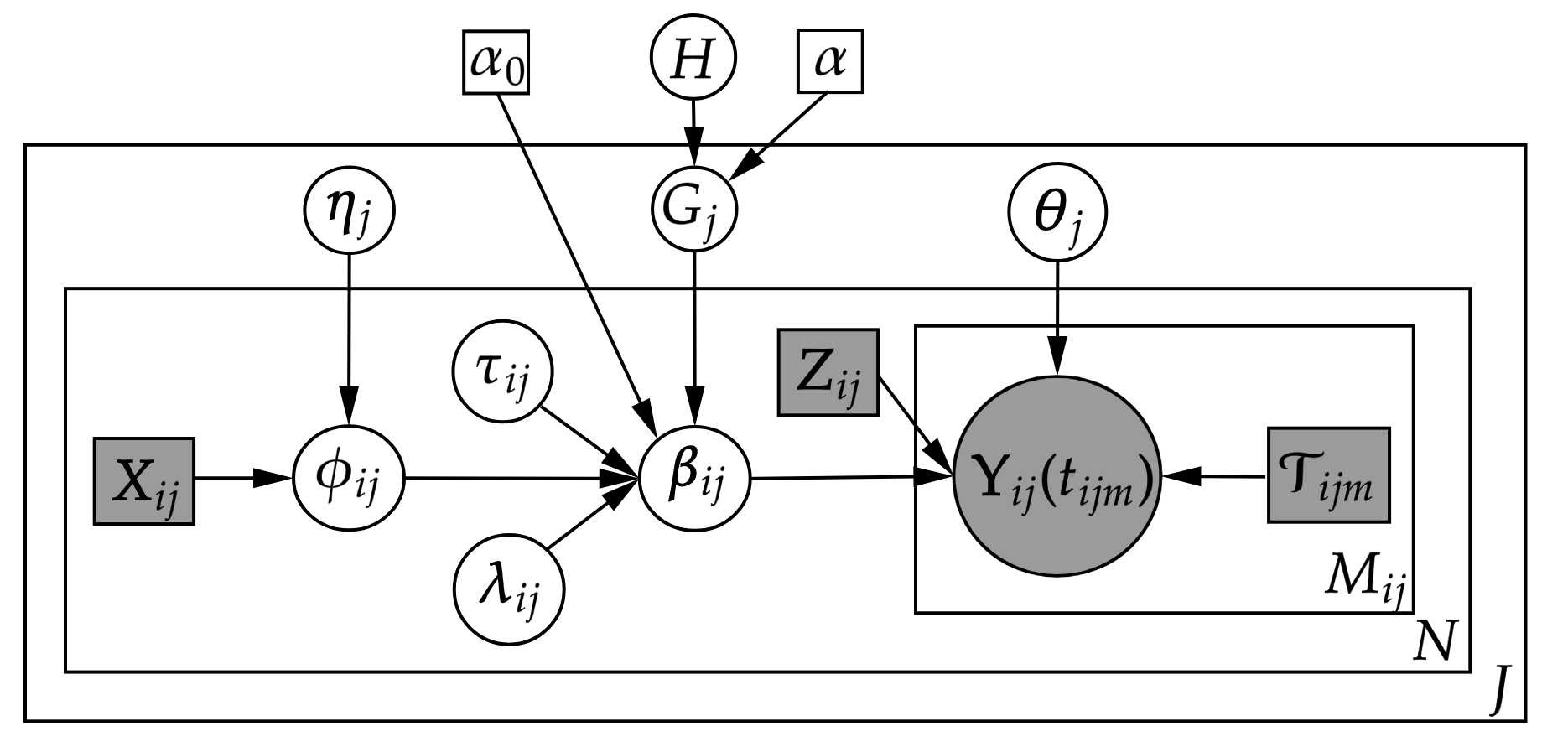}
\caption{A graphical representation of our hierarchical model 
} 
\label{fig:graphical}
\end{figure}

By introducing an additional parameter, $d_{ij}$, to denote the global cluster assignment for participant $i$ during period $j$, we can represent our full model in the following hierarchical structure:. 
\begin{equation}
\label{eq:hierarchical}
\begin{aligned}
\text{logit}\bigg(P(Y_{ij}(t_{ijm})=1|\bm{\beta}^*,d_{ij},\bm Z_{ij},\bm{\theta}_j)\bigg) &= \bm{\mathcal{T}}_{ijm}\bm{\beta}^{*}_{d_{ij}} + \bm{Z}_{ij} \bm\theta_j \\
\bm{\beta}^{*}_{d}|\tau^*_d, \bm\lambda^*_d &\stackrel{\text{ind}}{\sim} N(\bm{0}, \tau^*_d \bm\lambda^*_d) \\
\tau^*_d|\nu_{\tau_d}^* &\stackrel{\text{iid}}{\sim} \mathcal{IG}(1/2,1/\nu_{\tau_d}^*)\\
\lambda^*_{dq}|\nu_{\lambda_{dq}}^* &\stackrel{\text{iid}}{\sim} \mathcal{IG}(1/2,1/\nu_{\lambda_{dq}}^*)\\
\nu_{\tau_d}^* &\stackrel{\text{iid}}{\sim} \mathcal{IG}(1/2,1)\\
\nu_{\lambda_{dq}}^* &\stackrel{\text{iid}}{\sim} \mathcal{IG}(1/2,1)\\
\bm\theta_j &\stackrel{\text{iid}}{\sim} N(\bm 0, \bm {\Sigma}_\theta)\\
\{d_{11},\dots,d_{1N},\dots,d_{J1},\dots,d_{JN}\} &\sim \text{htRPM}(\phi,\alpha,\alpha_0) \\
\text{logit}(\phi_{ij}) &= \bm{X}_{ij} \bm{\eta}_j \\ 
\bm{\eta}_j &\stackrel{\text{iid}}{\sim} N(\bm\mu_\eta,\bm\Sigma_\eta),
\end{aligned}
\end{equation}
where $\bm{\mathcal{T}}_{ijm}$ denotes the $m^{th}$ row of matrix $\bm{\mathcal{T}}_{ij}$. A graphical representation of this hierarchical structure is provided in Figure \ref{fig:graphical}.

\color{black}
\subsection{Posterior Inference}
\label{sec:post}
For posterior inference, we implement a Gibbs sampling scheme. Briefly, we use the P\'olya-Gamma (PG) data augmentation technique of \cite{polson2013bayesian} to efficiently update the spline coefficients $\bm{\beta}_{ij}$ and the baseline effect coefficients $\bm{\theta}_j$ by introducing iid latent variables ${\omega_\beta}_{ijt}$ and ${\omega_\theta}_{ijt}$, which both follow a $\text{PG}(1,0)$ for each observation $Y_{ij}(t_{ijm})$. A similar technique is also used to update $\bm{\eta}_j$ for each $\gamma_{ij}$, together with latent variables ${\omega_\eta}_{ij}$. 
We follow 
\cite{page2021dependent} to update the auxiliary parameter $\gamma$. Next, the hierarchical clustering assignments are updated following the proposed Gibbs sampler in \cite{bassetti2020hierarchical}. Both steps are similar in spirit to algorithm 8 in \cite{neal2000markov}, where auxiliary variable procedures are applied.
Lastly, the horseshoe hyperparameters are updated following \cite{makalic2015simple}. We summarize our sampling scheme in Algorithm \ref{alg:mcmc} below and give full details on the parameter updates in the Supplementary Material.

\begin{algorithm}[htb!]
\caption{MCMC sampler }
\label{alg:mcmc}
\begin{algorithmic}[1]
\State Input data $\boldsymbol{Y}$, $\boldsymbol{X}$, $\boldsymbol{Z}$ and $\boldsymbol{\mathcal{T}}$
\State Initiate $\boldsymbol{\beta}$, $\boldsymbol{\theta}$, 
$\boldsymbol{d}$,
$\boldsymbol{\omega}$,
$\boldsymbol{\lambda}$, $\boldsymbol{\tau}$, 
$\boldsymbol{\nu}$,
$\boldsymbol{\theta}$,
$\boldsymbol{\gamma}$,
$\boldsymbol{\eta}$ 
\For{iteration $j = 1, \dots, J$}
    \For{iteration $i = 1, \dots, N$}
    \State Update $\gamma_{ij}$ following Equation (7) and (8) in \cite{page2021dependent}
    \EndFor
    \For{iteration $i = 1, \dots, N$}
    \State Update cluster assignments following the sampler in \cite{bassetti2020hierarchical}
    \EndFor
    \For{iteration $d = 1, \dots, D$}
    \State Update $\bm\beta^*_d$ with a Gibbs step following the P\'olya-gamma augmentation strategy \citep{polson2013bayesian}
    \State Update ${\boldsymbol{\lambda}}^*_d$, ${\tau^*_d}$, $\boldsymbol{\nu}_{\lambda^*_d}$,${\nu}_{\tau^*_d}$ with a Gibbs step \citep{makalic2015simple}
    \EndFor
    \State Update $\boldsymbol{\theta}_j$ with a Gibbs step
    \State Update $\boldsymbol{\eta}_j$ with a Gibbs step
\EndFor
\end{algorithmic}
\end{algorithm}

After burn-in and thinning, the remaining samples obtained from the MCMC algorithm are used for inference. To determine participants' cluster assignments for each smooth function, we use the sequentially-allocated latent structure optimization (SALSO) method to minimize the lower bound of the variation of information loss \citep{meilua2003comparing,dahl2021search}. \cite{dahl2021search} demonstrate the superior clustering performance of the SALSO method compared to alternative approaches. 
Note that our model requires predefined hyperparameters $\alpha$ and $\alpha_0$. To determine appropriate hyperparameters, we can use the shape of estimated trajectories within each cluster and consider the cluster size. If clusters have similar estimated trajectories or if some cluster sizes are small, it is recommended to choose lower values for the hyperparameters to reduce the number of clusters. 
We compare models using the Watanabe-Akaike information criterion (WAIC)  \citep{watanabe2010asymptotic}. Following \cite{gelman2014understanding},  WAIC values are calculated as $-2\times(\text{lppd} - p_{\text{WAIC}})$, where lppd represents the log pointwise predictive density and $p_{\text{WAIC}}$ estimates the effective number of parameters and penalizes complex models. These two values are calculated as
\begin{equation} 
\label{eq:waic}
\begin{aligned}
\text{lppd} &= \sum_{j=1}^J \sum_{i=1}^N \text{log}\bigg(\frac{1}{S}\sum_{s=1}^S P(Y_{ij}|\mathcal{B}^s_{ij},\bm{\theta}^s_j) \bigg) \\
p_{\text{WAIC}} &= \sum_{j=1}^J \sum_{i=1}^N \text{Var}_{s=1}^S \text{log}\bigg(P(Y_{ij}|\mathcal{B}^s_{ij},\bm{\theta}^s_j)\bigg),
\end{aligned}
\end{equation} 
where $\text{Var}_{s=1}^S$ represents the sample variance, and the superscript $s$ represents the $s^{th}$ draws from the posterior samples. In this study, we set $S$ equal to 10\% of the posterior sample size after burn-in.

It is important to note that our model is subject to label switching, as the partial likelihood from the auxiliary parameters $\gamma_{ij}$ is invariant when $\phi_{ij}$ are similar among participants.
For instance, if participant 1 and 2 are clustered together at period $j-1$ and separated at period $j$, then the joint probability of $\gamma_{1j} = 1$ and $\gamma_{2j} = 0$ would be same as the joint probability of $\gamma_{1j} = 0$ and $\gamma_{2j} = 1$ if $\phi_{1j}=\phi_{2j}$.
One potential solution to handle this issue is to incorporate participant-specific effects and/or covariates in $\phi_{ij}$ to prevent similar values across participants.

\section{Simulation}
\label{s:sim}
In this section, we evaluate the clustering and estimation performance of our proposed method on simulated data. 
As there are no other existing methods that explicitly model temporal functional clustering, we compare our proposed model (htRPM) 
to three other naive versions of the model. The proposed model differs from the naive models by incorporating hierarchical structure and accommodating temporal dependence, while the naive models lack one or both of these features. Specifically, we compare the model to a similar version of htRMP that does not allow for clusters to be shared between periods, tRPM, a model that assumes a HDP for the functional trajectories, HDP, and a standard DP for the functional trajectories, DP.  
By comparing these models, we aim to demonstrate the potential benefits of using temporal and hierarchical structures in time-correlated functional clustering for improving clustering and estimation performance. 

We explore the models in two simulated scenarios; one with independent clusters between periods and the other with varying degrees of temporal dependence between the clusters. In each scenario, we applied our proposed model to 50 replicated data sets. In each simulation, the MCMC algorithm was run for 5,000 iterations, 
where the first 3,000 iterations were treated as burn-in and the remaining 2,000 iterations were thinned to every 10$^{th}$ iteration, resulting in 200 iterations used for inference. 
Convergence of the models was determined using traceplots of the total number of global unique realizations, and traceplots of the coefficient estimates (see Supplementary Material for more details).
For all models, we assumed the prior of the global effect coefficients $\bm \theta_j \sim \text{N}(\bm 0, I)$, where $I$ indicates the identity matrix.
For all models, we set $\alpha=0.1$ to avoid potentially overestimating the number of clusters. For hierarchical models, we set  $\alpha_0=1$. All regression coefficients were initialized at 0, and all participants were assumed to be in the same cluster. 
Convergence of the models was determined using traceplots of the total number of unique clusters in the model and traceplots of the coefficient estimates, which can be found in the Supporting Information.

To determine the cluster allocation for each participant-specific smooth function across periods, we applied the SALSO method to the posterior samples, as described in section \ref{sec:post}.
Note that for non-hierarchical models (i.e., tRPM and DP), clusters were estimated individually for each period, whereas for hierarchical models (i.e., htRPM and HDP) cluster estimation includes all participants across all periods.
We then evaluated the clustering performance based on the variation of information (VI) \citep{meilua2007comparing} and the adjusted Rand index (ARI) \citep{hubert1985comparing}, which both measure closeness between the true and estimated clustering allocations. 
VI is always non-negative and values closer to 0 indicate better clustering performance. 
ARI takes on values between 0 and 1, with 0 indicating that two allocations do not agree on any pair of items and 1 indicating that the allocations are the same. 
Lastly, we evaluated the models' ability to recover the true smooth functions by measuring the mean squared error (MSE$_\mathcal{B}$) between estimated and true values for smooth functions at each observation time, averaged across participants and periods. That is,
\begin{equation*}
    MSE_\mathcal{B} = \frac{1}{\sum_{i=1}^N \sum_{j=1}^J M_{ij} } \sum_{i=1}^N  \sum_{j=1}^{J}  \sum_{m=1}^{M_{ij}} \bigg( \hat{\mathcal{B}}_{ij}(t_{ijm}) - \mathcal{B}_{ij}(t_{ijm})
    \bigg)^2,
\end{equation*}
where $\hat{\mathcal{B}}_{ij}(\cdot)$ is the estimated smooth function. A lower MSE$_\mathcal{B}$ indicates better estimation of the smooth function.

\subsection{Simulation 1: Data with Independent Clusters}
\label{sec:sim1}

\begin{figure}[]
\hspace*{-0.25in}
\centering
\begin{subfigure}[b]{0.5\textwidth}
     \centering
     \includegraphics[width=\textwidth]{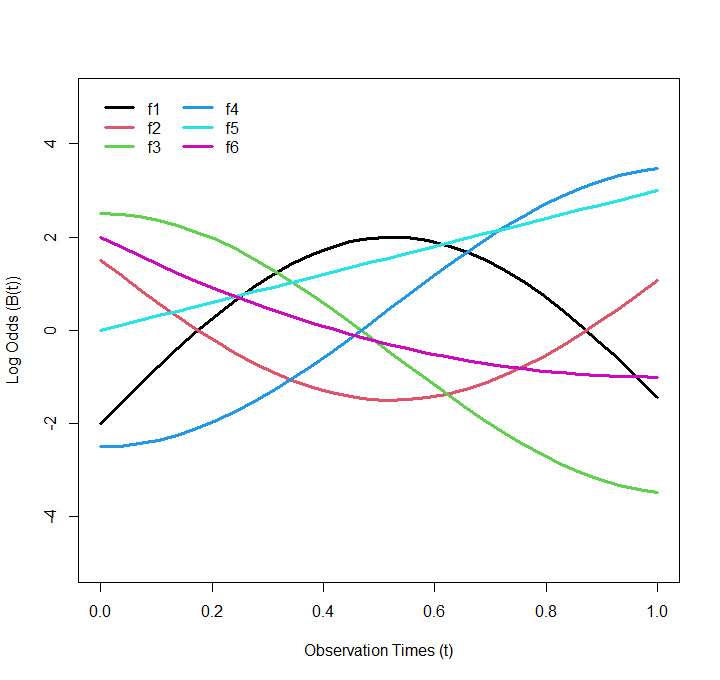}
\end{subfigure}
\caption{{\bf Simulation study:} Smooth functions used as functional random effects.} 
\label{fig:sf}
\end{figure}

In the first scenario, we evaluate the clustering and parameter estimation performance when the true partitions are independent between periods. 
We simulated data by setting the number of participants and periods to $N=50$ and $J=5$, respectively, similar to the application data. For each participant within each period, we generated $M_{ij} = 30$ observations at times $t_{ijm}$. Each observation time was sampled from a Uniform(0,1) distribution without loss of generality. Participant-specific smooth functions for each time period were randomly assigned within and across each period from the following set of functional trends: 
\begin{itemize}
    \item $f_1(t_{ijm}) = 4\text{sin}(3 t_{ijm}) - 2 $
    \item $f_2(t_{ijm}) = -3\text{sin}(3 t_{ijm}) +1.5$
    \item $f_3(t_{ijm}) = 3\text{cos}(3 t_{ijm}) - 0.5$
    \item $f_4(t_{ijm}) = -3\text{cos}(3 t_{ijm}) + 0.5$.
\end{itemize}
See Figure \ref{fig:sf} for a plot of these smooth functions. Two baseline covariates $\bm Z_{ij}$ were simulated from a standard multivariate normal distribution. The corresponding regression coefficients, including an intercept term, were set as $\bm\theta_j = (0.5,-0.5,0.3)^\intercal$. Note that the intercept terms in $\bm Z_{ij}$ can be considered as period-specific intercepts, whereas intercept terms in the smooth functions can be considered as participant-specific intercepts for the random effects. 

Table \ref{tab:siminp} displays the clustering and estimation performance for the four models. Overall, the HDP model obtained the best clustering and estimation performance. However, the proposed htRPM model, which unnecessarily accommodates potential temporal structure, performed similarly. We observed that the models with hierarchical processes (i.e., htRPM and HDP), which borrow information across periods, performed better than the non-hierarchical methods (i.e., tRPM and DP) in this setting. 

\begin{table}[tb]
\centering
\begin{tabular}{ccccc}
\hline
Model & VI & ARI & $MSE_\mathcal{B}$  \\ \hline
htRPM  
&  0.155 (0.150)  &  0.958 (0.041)  &  0.090 (0.024)  \\
HDP
&  0.116 (0.146)  &  0.969 (0.039)  &  0.079 (0.025)  \\
tRPM  
&  0.279 (0.209)  &  0.922 (0.061)  &  0.311 (0.240)  \\
DP 
&  0.152 (0.166)  &  0.959 (0.056)  &  0.106 (0.083)  \\ \hline
\end{tabular}
 \caption{{\bf Simulation study:} Means (standard deviation, SD) of the clustering and smooth function estimation performance for the four models across 50 replicated data sets, with true clustering assignments independent across periods. }
 \label{tab:siminp}
\end{table}

\subsection{Simulation 2: Data with Temporally Dependent Clusters}
\label{sec:sim2}

In the second simulation, we investigate the performance of the models when true clustering assignment for participants between periods $j$ and $j+1$ are correlated with varying degrees. 
To demonstrate that the flexibility in estimating the temporal dependence provides benefits in cluster and trajectory estimation, we also compare our proposed model to a simplified version of the htRPM model (htRPM$_\text{o}$), where the strength of the temporal dependence $\phi_{ij}$ is modeled by only an intercept term.
In this simulation, participants were initially allocated ($j=1$) to one of two clusters (i.e., smooth functions $f_1$ or $f_2$) with equal probability.
To generate temporal correlation across time periods with respect to the clustering allocation, we first determined whether or not a participant was able to change clusters from period $j$ to $j+1$ following the construction of $\gamma_{ij}$ in section \ref{sec:spline} 
, with parameters $\bm\eta_j$ and covariates $\bm X_{ij}$. 
We simulated $\bm\eta_j$ from $N(\bm\mu_\eta, 0.5 \bm I)$, where each element of the mean $\bm\mu_\eta$ was set as a sequence of $(-3,-2,-1,0,1,2,3)$ to represent 5\%, 12\%, 27\%, 50\%, 73\%, 88\% and 95\%  of participants fixed between the clustering transition from period $j$ to $j+1$.  
We generated three covariates for each participant, $\bm X_{ij}$, from $N(\bm 0, 0.5\bm I)$, and sampled $\gamma_{ij}\sim \text{Bernoulli}(\phi_{ij})$, where $\phi_{ij}=\text{logit}^{-1}(\bm X_{ij} \bm\eta_j)$.
Given $\gamma_{ij}=0$, participants were reallocated to any existing cluster or a new cluster with weights proportional to the sizes of the existing clusters or the concentration parameter.
Lastly, for each participant we generated three baseline covariates $\bm Z_{ij}$ from $N(\bm 0, 0.5\bm I)$. The corresponding regression coefficients were $\bm\theta_j = (0.5,-0.5,0.3)^\intercal$, including a population-level intercept term. 
In this setting, we added two more functional random effects to accommodate the creation of potential new clusters beyond $f_1$ to $f_4$:
\begin{itemize}
    \item $f_5(t_{ijm}) = 3 t_{ijm}$
    \item $f_6(t_{ijm}) = 3(t_{ijm}-1)^2 - 1$.
\end{itemize}
All prior settings were the same as described in section \ref{sec:sim1}, except we now assume the prior of $\bm\eta_j$ follows a N($\bm 0$, $5\bm{I}$).

In addition to evaluating clustering and estimation performance, we also investigated the ability of the models to recover indicators $\gamma_{ij}$ and the estimation accuracy of parameters $\bm\eta_{j}$. We evaluated the performance of $\gamma_{ij}$ by comparing the estimated $\gamma_{ij}$ at each iteration of the samples to the truth.

\begin{figure}[]
\hspace*{0.35in}
\centering
\begin{subfigure}[b]{\textwidth}
     \includegraphics[width=0.45\textwidth]{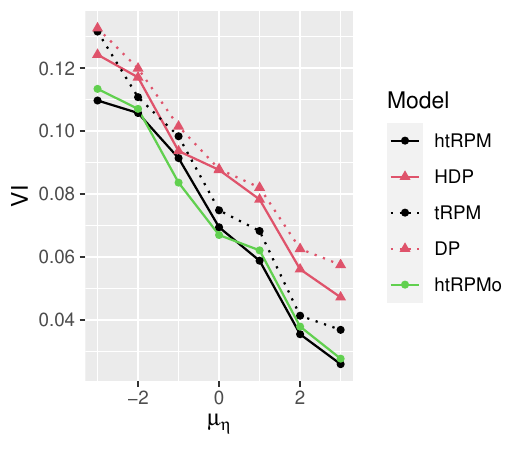}
     \includegraphics[width=0.45\textwidth]{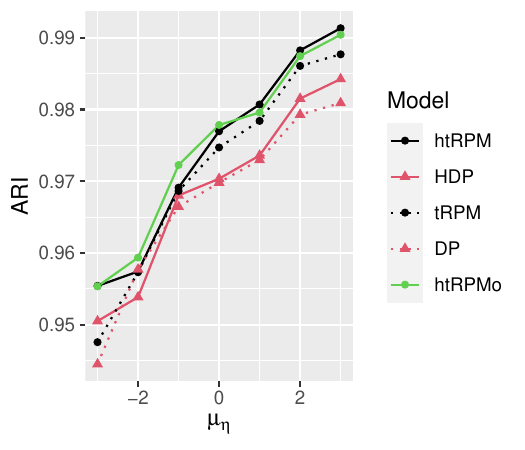}

     \includegraphics[width=0.45\textwidth]{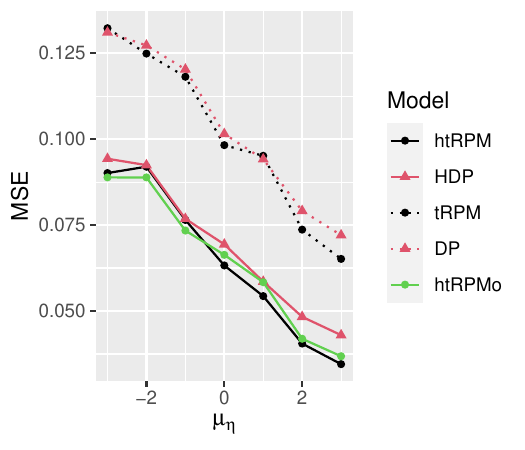}
     \includegraphics[width=0.45\textwidth]{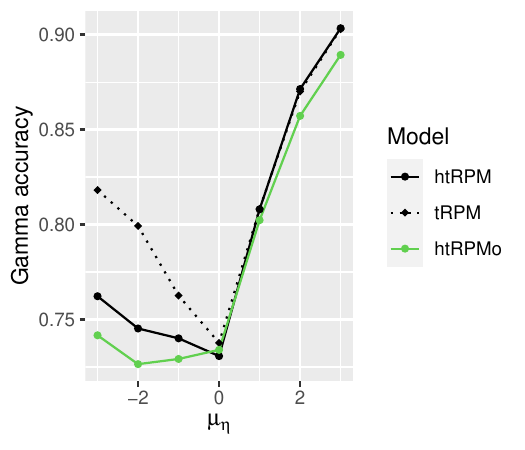}
\end{subfigure}
\caption{{\bf Simulation study:} Clustering and estimation performance of the five models. 
Clustering performance is measured by VI (lower indicates better performance) and ARI (higher indicates better performance). 
Trajectory estimation performance is measured by MSE. Performance of estimating the auxiliary parameter $\gamma_{ij}$ is measured by accuracy. Models are indicated by colors and line types: black solid lines - htRPM, red solid lines - HDP, black dotted lines - tRPM, red dotted lines - DP, green solid lines - htRPM without using covariates to model temporal dependence.}  
\label{fig:vi_ari_mse}
\end{figure}

Figure \ref{fig:vi_ari_mse} presents the results of the second simulation scenario. Overall, we observed that accommodating temporal structure when present improves clustering performance.  
We also found that the performance of the models improved when the true clusters were more correlated, indicating the models performed better with more cluster allocation similarity across time periods. 
Additionally, the models with hierarchical nonparametric priors obtained better clustering performance compared to non-hierarchical models. This was expected as different periods shared similar smooth functions.
We also observed better clustering performance when modeling the temporal correlation with covariates in most situations. Performance diminished with smaller effect sizes, as expected. 

Figure \ref{fig:vi_ari_mse} captures the advantage of the temporal models in smooth function estimation. As a result of being more accurate in clustering estimation, the temporally dependent models better recover the true smooth functions compared to independent models. 
As expected, the hierarchical nonparametric models also obtained better smooth function estimation performance compared to the non-hierarchical nonparametric models.

Lastly, the performance of estimating parameters $\gamma_{ij}$, which are only present in models htRPM, htRPM$_\text{o}$ and tRPM, can be found in Figure \ref{fig:vi_ari_mse}.
The results suggest that our model was effective in estimating $\gamma_{ij}$, particularly when a high proportion of participants were fixed.
As expected, increased bias may be observed when a greater number of $\gamma_{ij}$ values are zero (with $\mu_\eta$ being non-positive). This is because the accuracy of estimating $\gamma_{ij}$ relies on the cluster assignment both before and after the transition. Some participants could remain in the same cluster while being permitted to move.
Furthermore, we found that the htRPM model without using covariates to model the temporal dependence had poorer estimation of $\gamma_{ij}$ in most situations, except when $\mu_\eta=0$. 
This could be due to the parameters $\bm \eta_j$ being very close to zero, which makes covariates $\bm X_{ij}$ irrelevant to the outcome $\gamma_{ij}$.

Finally, we comment on the computational time of different models. Introducing hierarchical priors into the model led to a roughly 20\% increase in computational time, while incorporating the temporal structure into the model had minimal impact on the computational time.

\subsection{Sensitivity Analysis}
\label{sec:sens}

In this section, we assess the sensitivity of the results to specification of global and local concentration parameters $\alpha_0$ and $\alpha$ in the htRPM and HDP models.
We set each of the parameters to default values, as described in the section \ref{sec:sim2}, and then evaluated the effect of manipulating one term at a time on the model performance.
For each of the model specifications, we present the results on the same 50 replicated data sets generated in the simulation study. Other MCMC settings were set the same as the simulation study. On average, the WAIC values across all models were similar using simulated data.

Results for the clustering (measured by the VI) as well as the trajectory estimation performance (measured by the MSE) are reported in the Supplementary Material.
We found that model performance was relatively robust when the across-period 
concentration parameter $\alpha_0$ increased to 10 or the within-period concentration parameter $\alpha$ increased to 1. We observed marginal sensitivity when the across-period concentration parameter $\alpha_0$ decreased to 0.1  or the within-period concentration parameter $\alpha$ decreased to 0.01. The htRPM model outperformed the HDP model in terms of clustering performance and smooth function estimation, especially when there was strong temporal dependence.

\section{Application}
\label{s:appl}


We apply our method to intensive longitudinal data collected in the PREVAIL II study 
  \citep{Kendzor2023plus}.
The PREVAIL II study is a randomized controlled trial designed to evaluate the efficacy of offering abstinence-contingent financial incentives 
  (contingency management, CM) plus tobacco cessation counseling and pharmacotherapy 
for smoking cessation among socioeconomically disadvantaged adults,  relative to usual care (UC) alone (counseling and pharmacotherapy). 
In this study,  demographic and behavioral information were collected at baseline. Smartphone assessments were collected over a five-week period from one week prior to a scheduled quit attempt to four weeks after. 
Throughout the assessment period, participants completed daily diaries and received four random EMAs   per day via  study provided smartphones.  Participants were asked about  their recent smoking behaviors and various risk factors that might be associated with their smoking behaviors. 

The primary goal of this study was to model participant-specific smoking behavioral trajectories for each period. 
To maintain consistency with the study design, which involved verifying smoking status biochemically on a weekly basis and conducting counseling sessions every week, each period included a one week time-span. This approach resulted in a dataset with five periods.
We defined the outcome as whether or not a participant reported smoking in the 4 hours before each assessment, capturing momentary smoking behaviors during waking hours. 

To account for temporal dependencies between clusters across periods, we incorporated all available baseline covariates into the matrix $\bm X$. These included age (continuous), sex (binary; \textit{male} or female), heaviness of smoking index (HSI, continuous), race (binary; \textit{white} or minoritized race), years of education (continuous), and treatment (binary; \textit{UC} or CM). Terms in italics are treated as the reference group when defining indicator variables. We also included a time-varying covariate of whether or not a participant had used marijuana in the past week (binary; \textit{no} or yes). 
At each assessment, a participant was asked to report their current smoking status and potential risk factors. Thus to maintain temporality, we modeled the temporal dependency between the behavioral transition from week $j$ to $j+1$ with potential risk factors collected in week $j$. Similar temporal assumptions are commonly made in smoking behavior research  \citep{shiffman1996first,minami2014relations,bolman2018predicting,koslovsky2018bayesian,koslovsky2020bayesian,liang2021bayesian,liang2023functional}.

The data analysis was performed on 77 participants with at least five completed assessments of smoking behaviors and one completed assessment of marijuana use per week. The median   out of 35 possible  smoking responses per week was 30 (interquartile range (IQR): 23-34). 
The average numbers of responses for weeks 1 to 5 were 20.1, 33.2, 30.6, 30.4, 29.5, and that over the entire 5-week time span was 28.8. For this analysis, we set the variance term $\Sigma_\eta$ to $5\bm{I}$ and concentration parameters $\alpha_0 = 0.1$ and $\alpha = 0.01$. This choice was based on both cluster sizes and WAIC values, as illustrated in Table \ref{tab:appsens}. 

For posterior inference, we initiated the MCMC algorithm with a null model (i.e., $\bm{\beta} = \bm{0}$ and all participants in the same cluster). 
We ran the MCMC algorithm for 10,000 iterations, where the first 5,000 iterations were treated as burn-in and the remaining 5,000 iterations were thinned to every 10$^{th}$ iteration, resulting in 500 iterations used for inference.
Model convergence and appropriate mixing of the posterior samples were assessed using traceplots of the regression coefficients and the total number of active terms in the model. 
Goodness-of-fit was assessed using posterior predictive checks, by comparing replicated data sets from the posterior predictive distribution of the model to the observed data as described in \cite{gelman2000diagnostic}. See the Supplementary Material for more details on convergence and goodness-of-fit assessment.
We evaluated clustering allocation using the SALSO method. 

\subsection{Results}
\label{sec::appresults}

\begin{figure}[htb!]
\makebox[\textwidth][c]{
\includegraphics[width=0.7\textwidth]{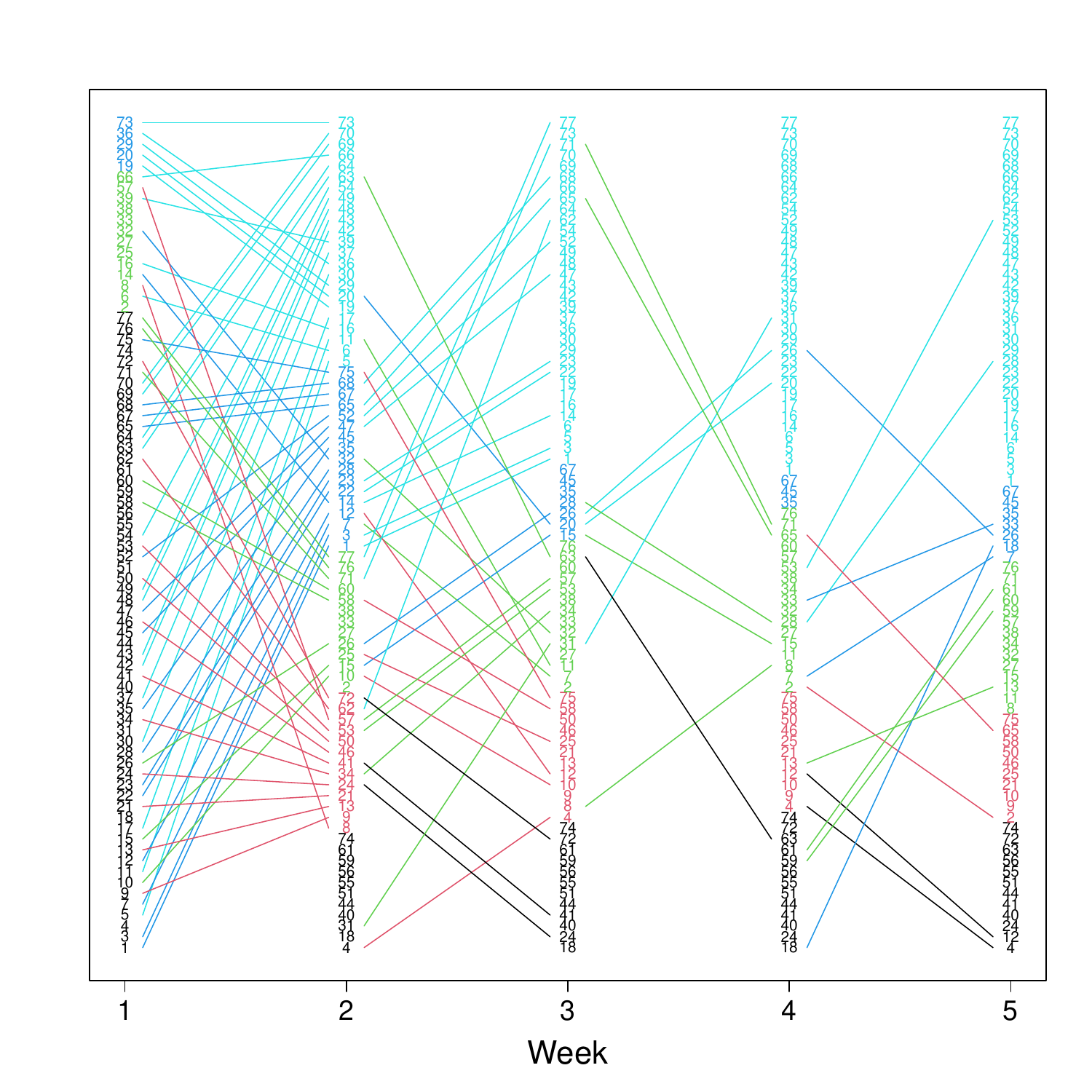}
\includegraphics[width=0.4\textwidth]{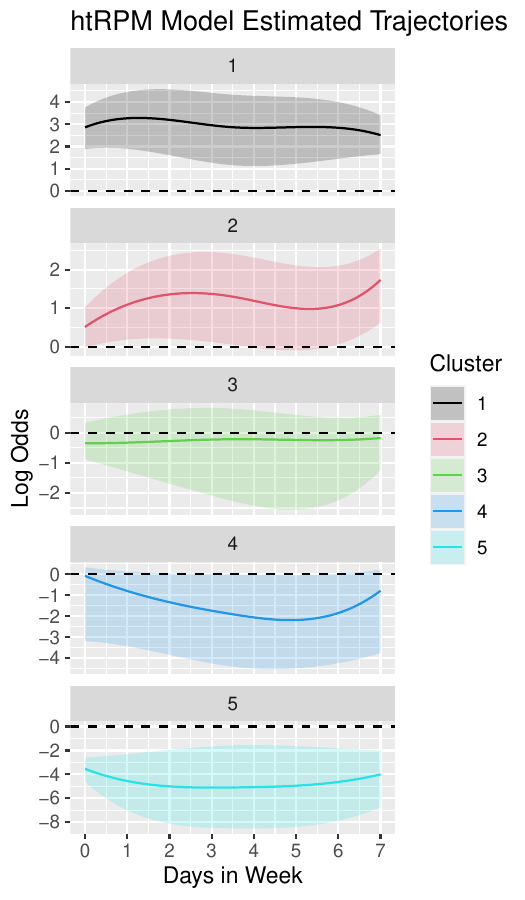}
}%
\caption{{\bf Case study:} Cluster allocation for each participant at each week estimated by the htRPM model. Different clusters are represented by different colors. Colored lines represent clusters participants moved to, if they did not stay in the same cluster.}
\label{fig:app_clus}
\end{figure}

The left subplot in Figure \ref{fig:app_clus} illustrates the cluster allocation evolution as estimated by our proposed model. 
Our model grouped participants by similar smoking behaviors on a weekly basis. Over the course of the study, it identified five clusters across all five periods. These clusters represent the specific smoking pattern to which a participant belongs for a given week. The lines from week to week indicate which cluster a participant transitioned to given they changed clusters. For instance, participant 8 was assigned to cluster 2 (red) in week 3 but transitioned to cluster 3 (green) in week 4, as indicated by the green line. The right subplot of Figure \ref{fig:app_clus} displays the five estimated functional trajectories of smoking, corresponding to the five identified clusters using the htRPM model. These clusters have been arranged in descending order of average log odds ratios for smoking across the week, labeled as cluster 1 (highest) through 5 (lowest). Specifically, the estimated mean for the log odds ratio for smoking for participants assigned to  cluster 1 ranged between 2 and 3. Cluster 2 had a log odds ratio for smoking slightly above zero. The corresponding log odds ratio for smoking for cluster 3 was around zero throughout the week. In cluster 4, the trajectory had a convex shape below zero, and cluster 5 exhibited a similar trend around a log odds ratio of -5. During the week before the quit attempt (week 1), we found the majority (77\%) of participants belonged to cluster 1 with the highest odds ratio for smoking throughout the week.  After the quit attempt, notable changes in cluster composition and smoking behaviors occurred. About 77\% of the participants moved to clusters with lower probabilities of smoking, and only 3\% moved to clusters with higher probabilities of smoking over time.  In comparison, transitions were less frequent from week 2 to 3, and the number of transitions were more balanced in both directions. Specifically, 25\% of the participants moved to clusters with lower odds of smoking, and 17\% moved to clusters with higher odds of smoking.  During the last two transitions, even fewer transitions occurred, where 88\% and 83\% participants stayed in the same cluster, respectively.

Prior research has demonstrated a dose-response relationship  between the intensity of certain clinical interventions for smoking (e.g., length of counseling sessions) and tobacco abstinence rates \citep{Fiore2008}.  
Identifying subgroups of participants who are more or less likely to smoke during a quit attempt from week to week enables prioritization of resources to those individuals who may need more support in order to stay quit. As patients in the current study attended weekly clinic visits, this model can help inform whether or not participants who are at a higher risk of smoking should receive additional resources (e.g., medications) in order to prevent transitions to a state with higher odds of smoking.

\subsection{Sensitivity Analysis}

We conducted a sensitivity analysis with varying across- and within-period concentration parameters 
to assess their impact on the application results. Given that the true outcomes are unknown in practice, we compared trajectory and cluster estimates across different parameter settings to those presented in Section \ref{sec::appresults}.

Based on our findings, our model consistently identifies similar major clusters when the concentration parameters are increased. The results, as shown in Table \ref{tab:appsens}, demonstrate that higher values of either or both concentration parameters resulted in a greater number of estimated clusters, as anticipated, where some of these clusters exhibited similar trajectories or were of smaller sizes. 
On the other hand, decreasing either $\alpha_0$ or $\alpha$ 
led the model to estimate one or two fewer clusters. However, even with these changes, the model still successfully categorized participants into groups with positive, around zero, and negative smoking log odds, which aligns with the main findings of our study.

\begin{table}[hbt!]
\centering
\begin{tabular}{lcll}
\hline
$(\alpha_0, \alpha)$ & \# Clusters & \multicolumn{1}{c}{Cluster Size} & \multicolumn{1}{c}{WAIC} \\
\hline
(1, 0.1)	& 9	& (1, 2, 9, 27, 36, 38, 42, 111, 119) & 6798 \\
(0.1, 0.1)	& 7	& (7, 25, 29, 50, 50, 103, 121) & 6833\\
(1, 0.01)	& 7	& (9, 19, 34, 40, 52, 112, 119) & 6837\\
\textbf{(0.1, 0.01)}& \textbf{5}& \textbf{(39, 46, 71, 107, 122)} & \textbf{7084}\\
(0.01, 0.01)	& 3	& (106, 139, 140) & 7485\\
(0.1, 0.001)	& 4	& (62, 67, 121, 135) & 7409\\
\hline
\end{tabular}
\caption{{\bf Case study:} Number of clusters, cluster sizes and WAIC estimated under different global and local concentration parameters. Models with lower WAIC values are generally preferred. Bold row indicates parameters used in the main application.}
\label{tab:appsens}
\end{table}

\section{Discussion}
\label{s:discuss} 
In this work, we have presented a fully Bayesian modeling approach for clustering functional data that accommodates temporal dependence among the clusters over time. Our model incorporates a hierarchical nonparametric prior to learn clusters within and between time periods. 
We provide additional flexibility in modeling the temporal dependence by incorporating covariates.  Through simulation, we have demonstrated that our proposed model outperforms independent clustering models with respect to clustering and estimation accuracy when temporal correlation between clusters exists. Our model also performs well even when true clusters are independent across periods. Furthermore, we have illustrated the practical application of our model by analyzing smoking cessation data. 

Our proposed model is designed for binary outcomes, as those collected in the application study. However, it can easily be adapted to accommodate continuous outcomes by replacing the logit link function with an identity link function. 
Similar to other Bayesian mixture models, our proposed method may encounter identifiability issues that can negatively impact inference. We have provided a simple solution for this issue in section \ref{sec:spline}. 
A known limitation of Dirichlet process mixture models is that they tend to overestimate the number of clusters in the data \citep{miller2013simple}. In the simulations and application study, we found that singleton clusters and clusters with very similar smooth function estimates may occur. To help protect against excess clusters, we recommend using the variation of information loss to provide posterior clustering estimates as it has shown to be more conservative with respect to the number of estimated cluster compared to alternative loss functions \citep{dahl2021search}. 
Further, we recommend performing thorough sensitivity analyses to understand the impact of prior specification on inference. Similar to existing algorithms for DP or HDP, another potential limitation of our model could be its inability to scale to large datasets. Addressing the scalability concern in future work may involve exploring variational inference \citep{hughes2015reliable} or parallel clustering \citep{meguelati2019dirichlet} solutions.





\section*{Code availability}
An \texttt{R} package, including code to simulate data, is available at \url{https://github.com/mliang4/ClusteringFunctionalTrajectoriesOverTime} (to be made public after acceptance of the paper).

\section*{Acknowledgments}
The PREVAIL II study was primarily supported by National Cancer Institute (NCI) grant R01CA197314 to Darla E. Kendzor. Additional support was provided by Oklahoma Tobacco Settlement Endowment Trust (TSET) grant R23-02 and NCI Cancer Center Support Grant P30CA225520 awarded to the Stephenson Cancer Center, the National Institute on Drug Abuse (R00DA046564 to Emily T. H\'{ebert}), and the Stephenson Cancer Center Mobile Health Shared Resource. This trial was registered at \url{ClinicalTrials.gov} (NCT02737566).

\bibliography{bibliography}

\end{document}